\documentclass[%
preprint,
amsmath,
amssymb,
showkeys,
]{revtex4-2}
\usepackage{amsmath}
\usepackage{graphicx}% Include figure files
\usepackage{dcolumn}% Align table columns on decimal point
\usepackage{bm}% bold  math
\usepackage{lineno}

\begin{document}
%\linenumbers 
%\preprint{APS/123-QED}

\title{All-optical photoacoustic tomography via beam deflection}

\author{Xingchi Yan$^\dagger$}
\email{xingchi\_yan@alumni.brown.edu (corresponding author)}
\affiliation{Department of Chemistry, Brown University, Providence, RI 02912, USA}
\affiliation{NSF-Simons Center for Mathematical and Statistical Analysis of Biology,
Harvard University, Cambridge, MA 02138, USA}

\author{Siyuan Song}
\thanks{These authors contributed equally to this work.}
\affiliation{School of Engineering, Brown University, Providence, RI 02912, USA}

\author{Hanxun Jin}
\affiliation{Department of Mechanical Engineering \& Materials Science,
Washington University in St. Louis, St. Louis, MO 63130, USA}
\affiliation{NSF Science and Technology Center for Engineering MechanoBiology,
Washington University in St. Louis, St. Louis, MO 63130, USA}

\begin{abstract}
Photoacoustic imaging (PAI) uniquely combines the advantages of optical contrast with deep tissue penetration capability of acoustic waves, enabling imaging at depths of several centimeters. Conventional photoacoustic imaging methods have relied on pulsed lasers to induce the photoacoustic effect, coupled with arrays of pressure transducers to detect the resulting ultrasound signals. In this work, we propose an alternative all-optical approach that leverages optical deflection to record photoacoustic waves by an array of detection beams. The measured signal is shown to be the Radon transform of the pressure gradients. An optimization-based inversion procedure is used to reconstruct the initial time pressure gradient field.  Subsequently,  a Galerkin method is used to reconstruct the pressure field from the pressure gradient field.  The new modality offers the potential for enhanced sensitivity and reduced signal distortion, advancing the capabilities of photoacoustic imaging beyond traditional transducer-based systems. 
\end{abstract}

\keywords{Multiphysics modality, Inverse problems,  All-optical  photoacoustic imaging, Radon transform, Finite element method}

%\keywords{Suggested keywords}%Use showkeys class optwas directedon if keyword
                              %display desid

\maketitle
%\tableofcontents

\section{Introduction}
\label{sec:introduction}
The photoacoustic effect, which combines strong optical contrast with deep tissue penetration of ultrasound, has shown great promise in bioimaging with applications ranging from imaging of the rat brain \cite{wang2003noninvasive}, tissues \cite{jathoul2015deep}, breast \cite{kruger2010photoacoustic} to the human brain \cite{na2022massively}. Owing to its distinctive properties, photoacoustic imaging offers high contrast for blood as well as excellent resolution, allowing a variety of applications in medicine, including tumor detection \cite{bouchard2014ultrasound, jiang2018photoacoustic, xu2006photoacoustic, wang2017photoacoustic}.

The conventional way for recording the emitted acoustic waves initiated by pulsed lasers to date is to use an array of pressure transducers. Different setups of the array of ultrasound transducers have been presented in \cite{bouchard2014ultrasound, jiang2018photoacoustic, kruger1995photoacoustic, kruger2017thermoacoustic}.  Several research groups use a frequency encoded, continuous laser beam to generate photoacoustic waves that are to be detected by conventional piezoelectric transducers \cite{mandelis1991progress}. The Fabry-Perot interferometer has been used to sense the pressure field generated by a scanning probe laser beam across the interferometer surface \cite{zhang2011multimodal, huynh2025fast}. A U-shaped array of fiber Fabry-Perot interferometers has been used to record the integrated pressure over the length of the fibers \cite{burgholzer2005thermoacoustic}. The detailed review for photoacoustic imaging can be found in Ref. \cite{jiang2018photoacoustic, wang2017photoacoustic}. 

The idea of beam deflection to record the photoacoustic effect has been reported in a few references \cite{johnson2016gas, maswadi2016all, yan2021abel, yan2022thesis}, including a photoacoustic microscopy based on beam deflection \cite{maswadi2016all}.  Optical recordings have three major advantages over traditional pressure transducers.  

First, pressure transducers, which are critical components in photoacoustic imaging systems, are far from ideal \cite{xiong2017photoacoustic, yan2024photoacoustic}. Thin film transducers such as PVDF have quite flat frequency response which is great for high-resolution imaging but suffer from low-sensitivity issues \cite{guggenheim2017ultrasensitive}. Crystals such as LiNbO$_3$ and solid transducers such as PZT have higher sensitivity than PVDF but suffer significantly distortion of actual photoacoustic pressure waveforms. The distortion comes from the mechanical resonances of the solid transductors that can significantly distort the measured photoacoustic pressure waveform \cite{li2009photoacoustic}. The existing pressure sensors constitute an inherent limitation in the measurement accuracy \cite{johnson2016gas}. Beam deflection signals recorded by photodiode detectors are not subject to the same distortion issues, as commercially available photodetectors used for data acquisition offer sufficient bandwidth to accurately capture the signal without distortion.

Secondly, the sensitivity of the piezoelectric transductors is fixed depending on the piezoelectric voltage coefficient ($p$C/Pa) and the noise in the pre-amplifier. In the modality introduced, the laser beam is bent by passing through the sound field, and by lengthening the distance between the pressure field and the position of the detectors, it is possible to have huge amplitude gain with only the pointing stability of the laser ultimately limiting the gain that can be obtained. The sensitivity of the laser beam deflection has been demonstrated in which the signals were obtained with a probe laser with an output of only 3mW \cite{barnes2014probe}.

Thirdly, the optical recording based on beam deflection has a wide bandwidth constrained only by the speed of the optical sensor and the ability of the electronics to read the voltage from a diode detector. The commercially available optical sensors can easily have response times on the order of 10 ns. 

Despite the exceptional sensitivity of optical beam deflection-based photoacoustic microscopy\cite{barnes2014probe}, its extension to tomographic imaging remains unexplored. In this work, we propose a new imaging modality that utilizes the recording of optical deflections to detect the acoustic fields. The models and numerical methods for the source identification of the modality are presented. Consider the optical setup as shown in Fig. \ref{fig:schematic} (a) where the body of interest lies above the $\Gamma$ plane. An array of probe beams arranges on the $\Gamma$ plane is directed onto a photodetector array designed to record deflections of the probe beams in the $x_3$ direction. Since the acoustic waves contain density of variations that perturb the index of refraction of the medium, the individual probe beams will be deflected at each point along their trajectory in the $x_3$ direction in an amount proportional to the density gradient in the photoacoustic wave. The overall deflection of each beam in the array will be proportional to the integral of the deflections along the beam path.  Probe beams and optical sensors can be arranged in different configurations along the domain boundary to extract partial information about the pressure gradient distribution.

Here, we aim to reconstruct the source -- the initial pressure fields and the underlying geometry -- from one configuration of the sensors. Due to the partial noisy information we get, the inverse problem is ill-conditioned, which can be iteratively solved via the adjoint operator of the inverse problem \cite{benning2018modern}. The algorithm involves formulating wave propagation with a specified trial source term, followed by corrections derived from applying an inverse or adjoint operator to the resulting error term. The pressure field can be numerically integrated from the pressure gradient
fields using finite element methods. 
The synthetic experimental sensor signal can be generated numerically via forward simulation of several known source terms that mimic experimental sources. The paper is organized as follows, the main setup of the experiment and the forward problem are shown in Section II. The inversion algorithm based on optimization and Galerkin methods is presented in Section III. The main results for simple and complex sources are shown in Section IV, and the discussion follows in Section V. 

\section{Physical Modeling}\label{sec:physics}
\subsection{Constitutive equations
}

\begin{figure}
    \centering
\includegraphics[width=\linewidth]{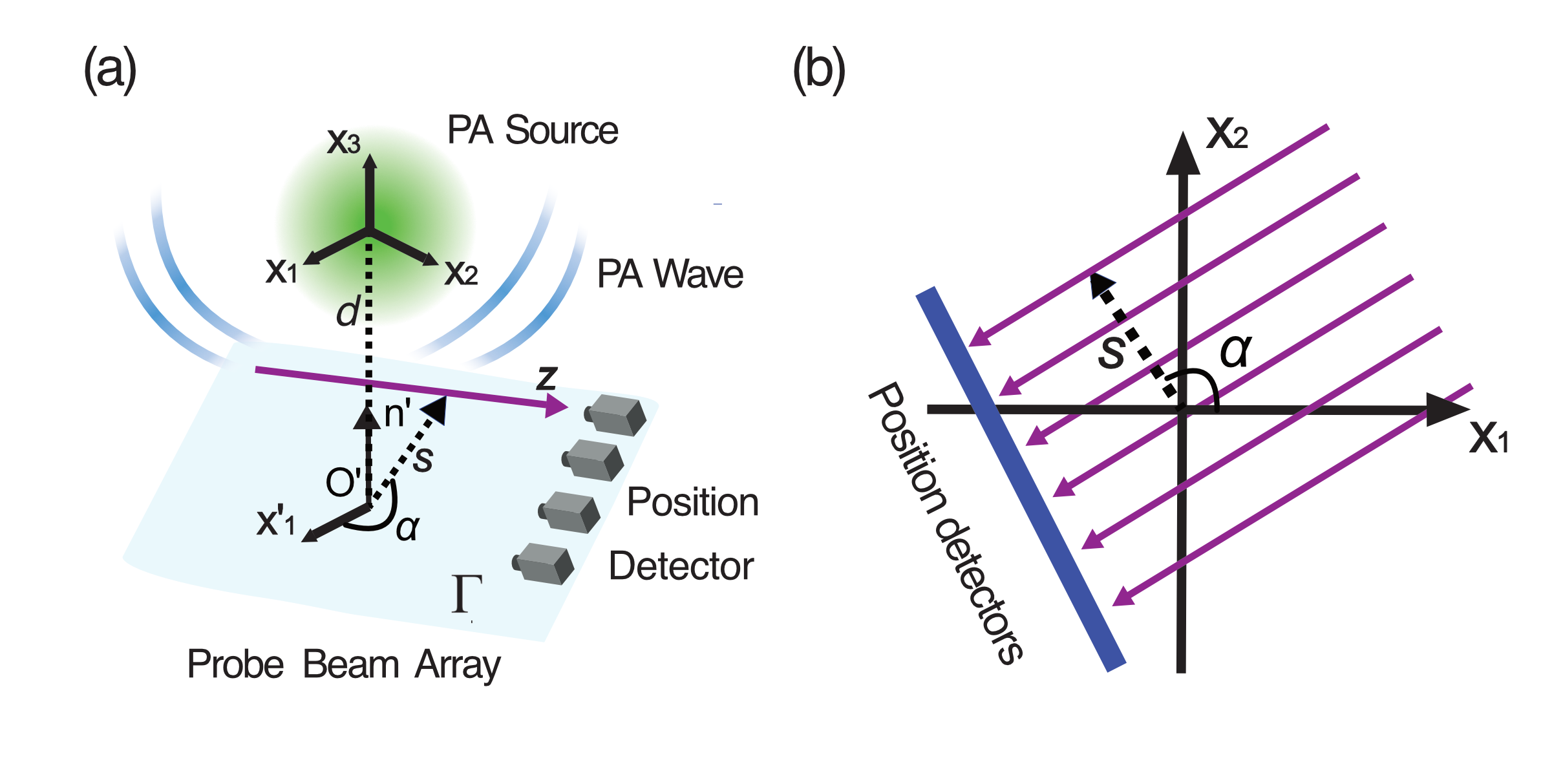}
    \caption{Schematic of the experimental setup for optical beam deflection tomography. (a) The absorber (shown in green) is irradiated by a short laser pulse and emits acoustic waves (shown in blue). As the acoustic waves propagate to the probe beam array plane $\Gamma$ (shown in light blue), the probing beam array (shown in purple) is deflected with the deflection angle along the $z$-axis recorded by an array of position detectors. The experiment can be repeated with the probe beam array rotated by an incremental angles $\alpha$ on the plane $\Gamma$ for each pulsed excitation. For reconstruction of the image, full rotation through 180 degrees must be recorded. Only one probe laser beam and a detector array are shown. 
    (b) A top view of the probe beam array and detectors. 
}
    \label{fig:schematic}
\end{figure}

Laser excitation generates acoustic waves propagating outwards, as illustrated in Figure ~\ref{fig:schematic}. Wave propagation in a homogenous compressible fluid under small deformation can be described via a system of linear equations based on mass conservation and linearized equations of states. For a small displacement of the particle in an ideal fluid, the conservation of the momentum is,
\begin{equation}
-\rho_0 \frac{\partial^2}{\partial t^2} \mathbf{u}(\mathbf{x}, t) = \nabla p(\mathbf{x})
     \label{eq}
\end{equation}
where \( \mathbf{u}(\mathbf{x},t) \) is the displacement of the particle, $t$ is the time, $\mathbf{x}$ is the global spatial coordinate. \( p(\mathbf{x}) \) is the pressure field, \( \rho_0 \) is the equilibrium density. For small perturbations of ideal fluid, the linearized equation of states is,
\begin{equation}
p(\mathbf{x},t) = c^2 \left( \rho(\mathbf{x},t) - \rho_0 \right)
\end{equation}
where \( c \) is the sound speed, \( \rho(\mathbf{x},t) \) is the instantaneous density. For small perturbation, the mass conservation is,

\begin{equation}
\frac{\partial \rho(\mathbf{x},t)}{\partial t} + \rho_0 \nabla \mathbf{v}(\mathbf{x},t) = 0
\end{equation}
where \( \mathbf{v}(\mathbf{x},t) = \frac{\partial \mathbf{u}(\mathbf{x},t)}{\partial t} \) is the velocity field. By combining the three relations, one can obtain the classic wave equation \cite{westervelt1973laser, diebold1991photoacoustic, calasso2001photoacoustic} for the pressure field as,

\begin{equation}
\frac{\partial p(\mathbf{x},t)}{\partial t^2} - c^2 \nabla^2 p(\mathbf{x},t) = 0
\label{pres_equation}
\end{equation}
With the given initial condition \(  p(\mathbf{x},t = t_0) = p_0(\mathbf{x}) \) and the boundary conditions \( B\), one can determine the pressure field  \( p(\mathbf{x},t) = \mathcal{F}(c, B) p_0(\mathbf{x}) \), where \( \mathcal{F}(c, B) \) is the forward operator. According to the linear feature of the hyperbolic partial differential equation, the pressure gradient field \( \nabla p(\mathbf{x},t) \) also satisfies the equation \ref{pres_equation}. Finite difference or spectral methods can be used to solve the forward problem \cite{cox2007k, treeby2010modeling,treeby2010k}.

\subsection{Optical Sensing}
At a distance from the source, a probe beam positioned on the measurement plane \( \Gamma \) traverses the PA field and is directed onto a photodetector (PD). The PA-induced local density fluctuations perturb the refractive index of the medium, causing the probe beam to be deflected along its path. The magnitude of this deflection is proportional to the local density gradient within the PA field. The PD captures the deflected probe beam signal, which encodes partial information about the pressure field \( p(\mathbf{x}, t) \) generated by the PA source. By collecting PD signals \( \beta \) from multiple detectors placed at different positions and orientations, and at different acquisition times, it's possible to reconstruct the initial PA source through the solution of an inverse problem. Ideally, multiple parallel beams and detectors would be used. However, to reduce costs, a simplified setup using a single probe laser and detector that can be scanned along the boundary can be implemented. 

The relationship between the PD signal \( \beta \) and the PA field \( p(\mathbf{x}, t) \) is formulated as follows. We consider a PA source \( p_0(\mathbf{x}) \) located near the origin of a global Cartesian coordinate system, which generates a time-dependent pressure field \( p(\mathbf{x}, t) \). A measurement plane \( \Gamma \) is positioned at a distance from the source. This plane is characterized by its unit normal vector \( \mathbf{n}' \) and its normal distance \( d \) from the origin. A local polar coordinate system \( (s, \alpha) \) is constructed on \( \Gamma \), centered at the projection of the origin onto the plane, denoted as \( O' \). For each point specified by \( (s, \alpha) \), we define a tangential line that passes through this point and is perpendicular to the radial vector. This tangential line represents the orientation of the probe beam, which is aligned with the corresponding PD. Along this line, we establish a Frenet–Serret frame, with the local coordinate \( z \) defined along the beam’s propagation direction. The entry and exit positions of the probe beam (i.e., its endpoints) are denoted by \( z_{\rm{L}} \) and \( z_{\rm{U}} \), respectively. The relative signal amplitude detected by the PD is then given by the Radon transform \cite{radon2007determination}:
\begin{equation}
  \beta= M(\alpha, s;\, \mathbf{n}', d, \tau)\, p(\mathbf{x}, t) = \int_{z_{\rm{L}}}^{z_{\rm{U}}}
  \mathbf{n}' \cdot \nabla p\big( \mathbf{x}(z;\, \alpha, s, \mathbf{n}', d),\, \tau \big)\, dz
  \label{radon_formula}
\end{equation}

Here, the measurement operator \( M(\alpha, s;\, \mathbf{n}', d, \tau) \) consists of three sequential sub-operations:
\begin{equation}
  M(\alpha, s;\, \mathbf{n}', d, \tau)
  = R(\alpha, s)\, \chi(\mathbf{n}', d,\tau)\, \mathbf{n}' \cdot \nabla,
\end{equation}
where:
\begin{itemize}
  \item \( \mathbf{n}' \cdot 
  \nabla \) computes the directional derivative of the pressure field;
  \item \( \chi(\mathbf{n}', d, \tau) \) extracts the values on the measurement plane \( \Gamma \) at the time \(t=\tau\);
  \item \( R(\alpha, s) \) denotes a truncated Radon transform along the path from \( z_{\rm{L}} \) to \( z_{\rm{U}} \).
\end{itemize}
Specifically, for a scalar field \( f(x_a, x_b) \), where \( x_a \) and \( x_b \) represent local Cartesian coordinates defined on \( \Gamma \), the operator \( R(\alpha, s) \) is defined as:
\begin{equation}
R(\alpha, s)\, f(x_a, x_b) = \int_{z_{\rm{L}}}^{z_{\rm{U}}}f\Big( s \cos\alpha + z \sin\alpha, s \sin\alpha - z \cos\alpha \Big)\, dz
\end{equation}
where the integration is performed along a line perpendicular to the radial direction defined by the polar coordinate \( (\alpha, s) \) on the plane. 

The setup parameters \( \alpha, s,\, \mathbf{n}', d, \tau \) can be adjusted for different configurations of data acquisition set up. These variations can be achieved through several methods, such as:
\begin{itemize}
  \item Rotating the target about the axis defined by \( \mathbf{n}' \), which changes the angle \( \alpha \) without altering the measurement plane \( \Gamma \) or the sensor offset \( s \);
  \item Setting a sequence of acquisition times \( \tau \);
  \item Configuring a series of sensor arrays with different values of \( s \), \( \mathbf{n}' \), and \( d \).
\end{itemize}

For convenience, we denote a single measurement by a given PD at a given time as \( M\). Our goal is to reconstruct the PA source based on the collection \(\mathcal{B}=\{ \beta\}\) of all such measurements \( \mathcal{M}=\{ M\} \).

\section{Inverse Problem}\label{sec:inverse}
\subsection{Mathematical Formulation}
The associated time-reversal problem aims to reconstruct the initial pressure distribution \( p_0(\mathbf{x}) \) from the measured optical deflection data \(\mathcal{B}\). The forward wave-propagation problem is deterministic and well posed. However, the inverse problem is not: for a limited number of measurements \(\mathcal{M}\), recovering \(p_0(\mathbf{x})\) from the measured signals becomes ill-posed, and a unique back-propagation operator does not exist. We reformulate this as an optimization problem to find an approximate field \( \tilde{p}_0 \) that satisfies all available physical and measurement-based constraints. 
Several numerical methods, including backpropagation  \cite{xu2002exact, xu2005universal, burgholzer2007temporal, clifton2020inverse}, adjoint-operator methods \cite{arridge2016adjoint, treeby2010photoacoustic}, and machine-learning-based inversions \cite{antholzer2019deep, hauptmann2020deep, jin2023recent}, have been developed for conventional photoacoustic reconstruction. In this section, we introduce a reconstruction algorithm based on the adjoint operator and implement using an ISTA-type (Iterative Shrinkage-Thresholding Algorithm) scheme \cite{beck2009fast}. 

To quantify this data-consistency requirement, the sensor loss is defined as,
\begin{equation}
L^{\mathrm{Sensor}}(\tilde{p}_0) = \sum_{M \in \mathcal{M}} w(M) \left\| M \mathcal{F} \tilde{p}_0 - \beta(M) \right\|
\label{eq:old_mini}
\end{equation}
where $\tilde{p}_0$ denotes a trial approximation of the true initial pressure field within the domain
of interest, \( w(M) \) is the weight assigned to the measurement \( M \in \mathcal{M} \), and \( \left\| \cdot \right\| \) denotes the \( L_2 \)-norm. The summation is taken over the set of all measurements $ \mathcal{M} $. The inverse problem can be formulated as the following optimization problem, $\min\limits_{\tilde{p}_0} (L^{\mathrm{Sensor}})$. First, non-optimized sensor placements can yield low information measurements, making direct inversion an underdetermined problem. Second, the measurement process involves Radon transformations as well as differentiation and integration operators, a direct minimization of Eq. \ref{eq:old_mini} would therefore require performing Radon and inverse-Radon transforms repeatedly, along with high-order numerical differentiation and integration at each iteration, leading to large computational cost and numerical instability. To address these two challenges, we introduce a reconstruction protocol that integrates (1) a structured measurement strategy, (2) a preprocessing procedure that extracts directional components of the measured signal, and (3) a reconstruction framework tailored to these processed measurements.

\paragraph{Measurement strategy.} A complete measurement protocol requires specifying three essential components:
(i) the spatial placement of sensors, 
(ii) the temporal sampling scheme, and 
(iii) any active adjustments applied during data acquisition. 
In this work, these components are defined as follows.

\begin{itemize}
  \item 
  \textbf{Sensor Placement:} 
  The target is placed inside a cubic domain whose six boundary faces are 
  denoted by $\Gamma_{x_1^-}$, $\Gamma_{x_1^+}$, 
  $\Gamma_{x_2^-}$, $\Gamma_{x_2^+}$, 
  $\Gamma_{x_3^-}$, and $\Gamma_{x_3^+}$. 
  Measurement planes $\Gamma$ are restricted to these faces. 
  To recover pressure derivatives in all three coordinate directions 
  $x_1, x_2, x_3$, at least three mutually orthogonal faces must be instrumented, 
  with up to all six faces available. 
  On each selected face $\Gamma$, sensors are arranged such that their 
  in-plane offsets $s$ are uniformly distributed along a prescribed scanning 
  direction within the face.

  \item 
  \textbf{Acquisition Rate:} 
  All sensors collect data simultaneously using uniform time increments.

  \item 
  \textbf{Rotational Sampling:}
  For each measurement face $\Gamma$, the target and sensor array undergo relative rotation about the outward normal direction $\mathbf{n}'$ of the face. The rotation angles $\alpha$ are uniformly sampled within $0^\circ \le \alpha \le 180^\circ$.
\end{itemize}

\paragraph{Directional Preprocessing.}

The measurement strategy \( \mathcal{M} \) is divided into three mutually disjoint subsets, \( \mathcal{M} = \mathcal{M}_1 \cup \mathcal{M}_2 \cup \mathcal{M}_3 \), based on the normal vector direction of each measurement plane. For a given measurement plane $\Gamma$ and the time $\tau$, the ordered distribution of measurements with respect to \( \alpha \) and \( s \) allows the inverse Radon transform to recover the pressure gradient field in local Cartesian coordinates via \(R^{-1}\mathcal{B}_m(\Gamma,\tau)\), where \( \mathcal{B}_m(\Gamma,\tau) \) denotes the collection of PD signals associated with the subset \( \mathcal{M}_m \) corresponding to the given plane $\Gamma$ and the time $\tau$. Here, \(R^{-1}\mathcal{B}_m(\Gamma,\tau)\) represents the post-processed partial information extracted from the measurement data, capturing the pressure-gradient component (i.e., \(\partial p / \partial x_m\)) evaluated on a uniform Cartesian grid. This quantity serves as a refined reference signal compared with the raw measurements \(\mathcal{B}_m\).

In our setup, the photodetector measures the pressure-gradient field, whose Fourier-domain amplitude scales as $|k|p(k)$, where $k$ denotes the wavenumber. This leads to enhancement of high-frequency components and attenuation of low-frequency and DC components. Although deep-tissue signals are dominated by low frequencies due to biological attenuation, the gradient field $\nabla p$ encodes the spatial variations such as boundaries and interfaces -- that carry essential structural information. Consequently, the proposed gradient-based formulation preserves the diagnostically meaningful components while suppressing global backgrounds that are irrelevant to morphology. In future studies, combining the proposed modality with conventional techniques may yield enhanced performance through complementary frequency sensitivities and further improve image fidelity.

\paragraph{Reconstruction Framework.}{
Given the extracted pressure-gradient information \(R^{-1}\mathcal{B}_m(\Gamma,\tau)\), we reconstruct the directional derivatives \(\nabla_m \tilde{p}_0\) rather than the pressure field \(\tilde{p}_0\) itself. This choice avoids repeated numerical differentiation, integration, and Radon/inverse-Radon operations within each reconstruction iteration. For each orientation \(m \in \{1,2,3\}\), the forward problem to get the pressure gradient information in the given measurement plane $\Gamma$ and time $\tau$ is $\chi(\Gamma,\tau)\, \mathcal{F} \nabla_m \tilde{p}_0$. The data-consistency requirement is quantified through the following loss functional,}
\begin{equation}
    L^{\mathrm{Sensor\!-\!IR}}_{m}(\nabla_m \tilde{p}_0) =
    \sum_{\Gamma,\tau \in \mathcal{M}_m} 
    \Big\|
        \chi(\Gamma,\tau)\, \mathcal{F}\nabla_m \tilde{p}_0
        - R^{-1}\mathcal{B}_m(\Gamma,\tau)
    \Big\|.
\end{equation}
where the operator $\nabla_m$ denotes the directional derivative with respect to the $x_m$ axis. 
The summation runs over all measurement planes within \( \mathcal{M}_m \); depending on the setup, each \( \mathcal{M}_m \) may include one or two planes. The quantity $\chi(\Gamma,\tau)\, \mathcal{F}\nabla_m \tilde{p}_0$ represents the forward-simulated pressure gradient on the measurement plane $\Gamma$, while $R^{-1}\mathcal{B}_m(\Gamma,\tau)$ denotes the reconstructed gradient obtained from the inverse Radon transform of PD data. The optimization problem is reformulated as, 
\begin{equation}
    \min_{\nabla_m \tilde{p}_0} \Bigg( \lambda J(\nabla_m \tilde{p}_0) +
    L^{\mathrm{Sensor-IR}}_{m}(\nabla_m \tilde{p}_0) \Bigg)
    \label{eq:new_mini}
\end{equation}
The term \( \lambda J(\nabla_m \tilde{p}_0) \) introduces regularization to incorporate prior knowledge. Equation~(\ref{eq:new_mini}) defines three independent inverse problems—one per spatial direction—each solved separately.
Given the expected sparsity of the true pressure gradient field, \( J(\nabla_m \tilde{p}_0) \) is chosen as a \( L_1 \) norm regularization term to promote edge preservation and improve the recovery of the source topology. Each subproblem is solved using the iterative shrinkage-thresholding algorithm (ISTA) with the adjoint operator \( \mathcal{F}^* \) \cite{treeby2010photoacoustic, arridge2016accelerated}. Finally, the initial pressure field \( \tilde{p}_0 \) is reconstructed by solving the Poisson equation based on the full gradient field \( \nabla \tilde{p}_0 \) with reference pressure.

\begin{figure*}
    \centering
    \includegraphics[width=0.8\linewidth]{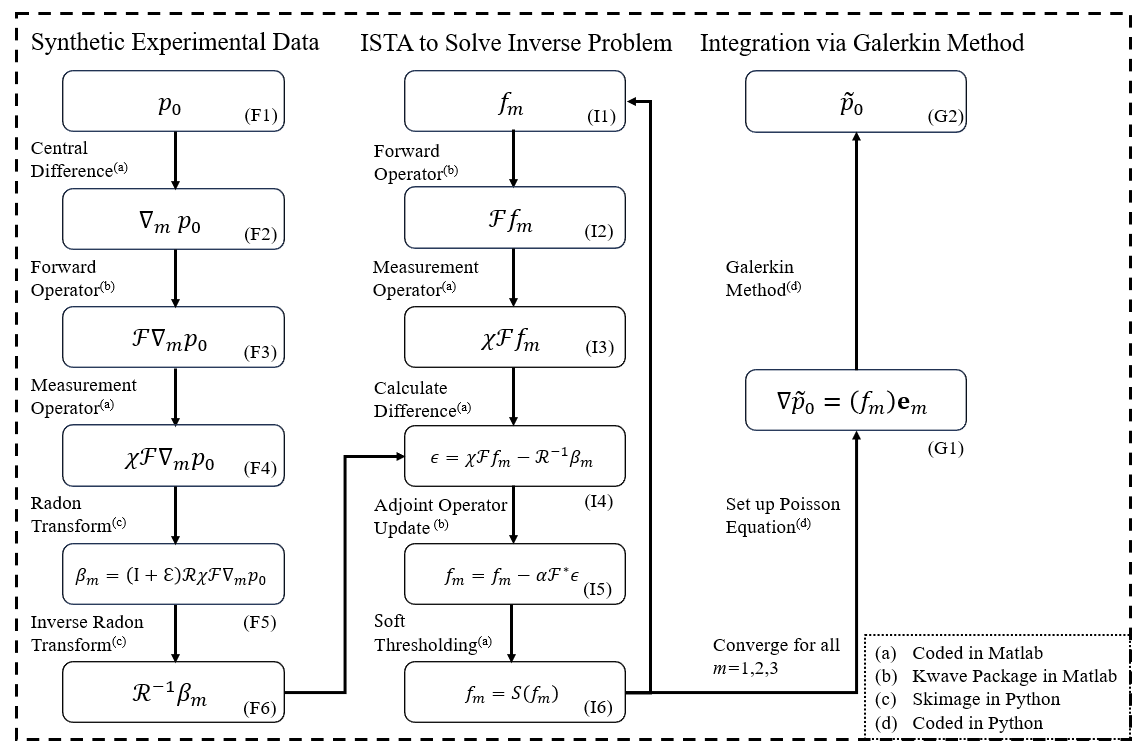}
    \caption{Algorithmic framework. The first column illustrates the forward simulation process used to generate synthetic experimental data. Starting from the bottom-left block, the six stages are: (F1) the initial pressure field, (F2) the corresponding pressure gradient field, (F3) the temporal evolution of the gradient field, (F4) its restriction to the measurement plane, (F5) the extraction of Radon transform values at sampled points, and (F6) the inverse Radon transform of the extracted data. The second and third columns depict the inverse reconstruction process. Beginning at the top-center, the seven stages include: (I1) the trial initial pressure gradient field, (I2) its simulated temporal evolution, (I3) the extraction of values at sampled points, (I4) the residual between trial and synthetic data, (I5) the adjoint operation to compute the correction term, (I6) the update of the trial gradient field using the correction, (G1) the formulation of the Poisson equation, and (G2) the final integration to reconstruct the initial pressure field.}
    \label{fig:algorithm}
\end{figure*}

\subsection{Numerical Algorithm}
The flowchart in Figure ~\ref{fig:algorithm} illustrates the complete algorithm to solve the minimization problems defined in equation~(\ref{eq:new_mini}). The algorithm consists of two major components:
(1) solving inverse problem using iterative methods, and  
(2) solution of the Poisson equation via the Galerkin method. 

In practical scenarios, the ground truth pressure source is unknown, and only the measured sensor data (corresponding to block F6) are available. In this study, we first generate synthetic experimental data using a known ground truth (block F1). The pressure gradient field in the direction \( \mathbf{x}_m \) is computed using a central difference scheme. To simulate the wave propagation, we employ the k-Wave package\cite{treeby2010k}, incorporating the constitutive relations introduced in the previous section. The measurement data are then extracted at predefined measurement planes (block F4). In practical experiments, optical sensors can measure pressure gradients in a way analogous to the Radon transform. To mimic this process, we apply the Radon transform and superimpose random noise (block F5). Synthetic noise follows a uniform distribution within the range \( (-\epsilon, \epsilon)|\beta| \), where \( |\beta| \) denotes the local absolute amplitude of the signal. The resulting data are then reconstructed using the inverse Radon transform to recover the pressure gradient distribution at discrete boundary points (block F6), completing the synthetic data generation process.

The second column of the flowchart represents the inverse reconstruction stage using ISTA. A trial pressure gradient field \( f_m = \frac{\partial \tilde{p}_0}{\partial x_m} \) is initialized. The simulation steps (I1) through (I3) mirror the forward process (F2) through (F4), yielding predicted sensor data. This data is compared with the synthetic measurements (block I4), and the residual is processed using the adjoint operator to generate a correction term (block I5). The correction is refined using a soft-thresholding operator derived from the \( L_1 \)-norm regularization:
\begin{equation}
\mathcal{S}(\cdot) = \mathrm{sign}(\cdot) \max\left( |\cdot| - \lambda, \, 0 \right),
\end{equation}
where the regularization parameter \(\lambda = 0.01\) is selected based on empirical optimization.

The final stage, shown in the third column, involves solving the Poisson equation to reconstruct the initial pressure field. The updated pressure gradient field at nodal points serves as the input. A finite element solution based on the Galerkin method is then employed to recover the underlying pressure source distribution.

\subsection{Numerical Implementation}
In this section, we briefly introduce the numerical implementation of the flow chart. Consider a cubic domain with the edge length of \( l = (N^{[F]} - 1) \Delta^{[F]} \), where
\( N^{[F]} = \left( 2n^{[F]} + 1 \right) \) is the number of nodes along the edge, \( \Delta^{[F]} \) is the interval size, the perfect match layer (PML) boundary conditions are applied to all six boundary surfaces. The space is discretized uniformly to \( N^{[F]} \times N^{[F]} \times N^{[F]} \) grid points \( x_{ijk} = (i,j,k) \Delta^{[F]} \), where \( i,j,k = -n^{[F]}, ..., -1, 0, 1, ... n^{[F]} \). To maintain numerical stability when solving the hyperbolic PDE, the CFL number requires,
\[
(\rm{CFL})^{[F]} = \frac{c}{\Delta^{[F]}/(dt)^{[F]}} < 1,
\]
where \( (dt)^{[F]} \) is the size of the time step, 
\(
N_t^{[F]} = \frac{\sqrt{3}l}{(dt)^{[F]}}
\)
is the maximum number of time steps ensuring that there is enough time for the wave propagating from one corner of the calculation domain to the opposite corner of the calculation domain. 

\begin{table}[h]
    \centering
    \caption{Spatial and temporal discretizations}
    \label{tab:parameters}
    \begin{tabular}{|c|c|c|}
        \hline
        \textbf{Parameter Name} & \textbf{Symbol} & \textbf{Value (Unit)} \\
        \hline
        Number of Nodes along the edge & \( N^{[F]} \) & 9, 17, 33, 65, 129 \\
        \hline
        Interval Size & \( \Delta^{[F]} \) & 1 mm \\
        \hline
        The number of nodes for the PML layer & \( N^{[F-PML]} \) & 10 \\
        \hline
        Sound Speed (in water) & \( c \) & 1500 m/s \\
        \hline
        CFL number & \( (CFL)^{[F]} \) & 0.25 \\
        \hline
        Number of time steps & \( N_t^{[F]} \) & \( \frac{\sqrt{3}l}{(dt)_{Forward}} \) \\
        \hline
        Learning rate in the gradient descent method & \( \nu \) & 0.1 \\
         \hline
        Regularization parameter & \(\lambda\) & 0.01
        \\
        \hline
    \end{tabular}
\end{table}

We propose a characteristic methodology for sensor placement that can be easily adapted to other strategies. Here we place the sensor at the position (1) 
\(
\left( -n^{[F]}, -n^{[F]} : \frac{n^{[F]}}{n^{[s]}} : n^{[F]}, -n^{[F]} \right) \Delta^{[F]},
\)
facing the \( x_3 \) direction to extract the information of \( \frac{\partial p}{\partial x_1} \);
(2) 
\(
\left( -n^{[F]}, -n^{[F]}, -n^{[F]} : \frac{n^{[F]}}{n^{[s]}} : n^{[F]} \right) \Delta^{[F]},
\)
facing the \( x_1 \) direction to extract the information of \( \frac{\partial p}{\partial x_2} \);
(3) 
\(
\left( -n^{[F]} : \frac{n^{[F]}}{n^{[s]}} : n^{[F]}, -n^{[F]}, -n^{[F]} \right) \Delta^{[F]},
\)
facing the \( x_2 \) direction to extract the information of \( \frac{\partial p}{\partial x_3} \).
Here \( n^{[s]} \) is a factor of \( n^{[F]} \). After the relative rotation of the source term, we could obtain a series of signals which are the Radon transform of the field information. For the integration outside of the calculation domain, the function value is considered zero due to the exponential decaying field in the viscous media. The Radon transform for the pressure gradient field in three directions, and \( N_t^{[F]} \) time steps, result in a matrix of dimension 
\(
3 \times N_t^{[F]} \times N^{[F]} \times N^{[\alpha]},
\)
where the third index refers to the distance of the sensor orientation to the surface center, the fourth index refers to a series of rotation angles. To simulate the measurement noise in the measurement, gaussian noise are added after the Radon transform of the pressure gradient fields.

Consider \( N^{[S]} = 2n^{[S]} + 1 \) sensors as introduced with the coarse-grained matrix with the dimension 
\(
3 \times N_t^{[F]} \times N^{[S]} \times N^{[\alpha]}.
\)
The inverse Radon transform gives the pressure gradient field on the boundary surface with the dimension 
\(
3 \times N_t^{[F]} \times N^{[S]} \times N^{[S]},
\)
defined in the new grids
\(
x_{ijk} = \left( i,j,k \right) \frac{n^{[F]}}{n^{[S]}} \Delta^{[F]},
\)
where 
\(
i, j, k = -n^{[s]}, ..., -1, 0, 1, ... n^{[s]}.
\) 
The Radon transform and the inverse Radon transform were implemented via the Python-Scikit-Image package \cite{van2014scikit}.
In the previous section, we obtain the synthetic pressure gradient field history 
\( F \nabla p_0 (x_{ijk}) \) 
at the preassigned boundary plane for all \( N_t^{[F]} \) time steps. The optimization problem in Equation \ref{eq:new_mini} can be solved as three independent problems with respect to three directions, via the adjoint operator \( F^* \) \cite{arridge2016adjoint}, as introduced in the second column of Figure ~\ref{fig:algorithm}. 

We start with an initial guess for the pressure-gradient field 
\( \nabla_m \tilde{p}_0 \)  as all zero
at \( t = t_0 \) and then calculated the evolution history 
\( \nabla_m \tilde{p} \) 
via the forward simulation. The calculation domain was discretized to 
\( N^{[I]} \times N^{[I]} \times N^{[I]} \) 
grid points, where 
\(
N^{[I]} = 2n^{[I]} + 1
\)
is the number of nodes along the edge direction. 
For convenience, we take 
\(
\frac{n^{[I]}}{n^{[S]}} \in \mathbb{N}
\)
in the current analysis to ensure that the observations are located at a subgroup of the grid points of the calculation domain. 
By comparing the guessed value of the pressure gradient at the boundary nodal points with the measurement, we will obtain the residue $\epsilon$. By applying the adjoint operator $F^*$ to the residue $\epsilon$, we got a correction term for the guessed value of the initial pressure gradient distribution. 

$\nabla_m \tilde{p}$ is iteratively updated based on the learning rate 0.1. The iterative trial-correction algorithm is implemented based on the adjoint operator and the ISTA algorithm, where the $L_1$ norm is chosen for the regularization term and the corresponding weight is chosen as 0.1. The whole optimization process includes 50 iterations. The resulting pressure gradient field is numerically integrated via the Galerkin approach based on Poisson’s equations. 

To evaluate the performance of the image reconstruction, we have defined the relative mean square error (RMSE), the peak signal-to-noise ratio (PSNR) and the structural similarity index measure (SSIM). The relative mean square error (RMSE) is defined as \cite{geman1992neural, wang2004image},

\[
e^{\rm{RMSE}} = \frac{\| f^{\rm{True}} - f^{\rm{Pred}} \|}{\| f^{\rm{True}} \|}
\]

where \( f^{\rm{True}} \) is the normalized ground truth of the field, \( f^{\rm{Pred}} \) is the normalized prediction based on image reconstruction, \(f=\frac{dp}{dx_1},\frac{dp}{dx_2},\frac{dp}{dx_3},p \). The peak signal to noise ratio (PSNR) and the structural similarity index measure (SSIM) are defined as \cite{geman1992neural, wang2004image},

\[
e^{\rm{PSNR}} = 10 \cdot \rm{log}_{\rm{10}} \left( \frac{(N^{[\rm{I}]})^3}{\left( \| f^{\rm{True}} - f^{\rm{Pred}} \| \right)^2} \right)
\]
\[
e^{\rm{SSIM}} = 
\frac{
    (2\mu^{\rm{Pred}} \mu^{\rm{True}} + 0.0001)(2\sigma^{\rm{Pred-True}} + 0.0009)
}{
((\mu^{\rm{Pred}})^2 + (\mu^{\rm{True}})^2 + 0.0001) \cdot ((\sigma^{\rm{Pred}})^2 + (\sigma^{\rm{True}})^2 + 0.0009)
}
\]
where \( \mu \) and \( \sigma \) denote the mean and standard deviation of \( f \), respectively. To simplify the analysis, we always take 
\( N^{[F]} = N^{[s]} = N^{[I]} \), indicating the number of nodes in the synthetic experimental data generation, the number of sensors along the edge, and the number of nodes in the inverse formulation are identical. The noise term is kept zero except for the parameter study section.

\section{Result}

Given the proposed imaging setup and the numerical challenges associated with the inverse problem, we begin by validating the approach through numerical simulations.  A simple setup involves determining the spatial and temporal profiles of the photoacoustic effect from an array of optically thin fluid spheres \cite{diebold1988photoacoustic, diebold1990photoacoustic, yan2021generation} irradiated by a Gaussian time profile laser pulse, from which the derivative of the acoustic field with respect to $x$, $y$ and $z$ can be determined and measured as described in Section \ref{sec:physics}. Given the flexibility of the numerical implementation, we opted for more general sources to enable broader applicability. Numerical convergence was shown in Section \ref{subsec:convergence} , followed by validation of the method using simulations of the \textit{Shepp3d} phantom, cubic fields, and mechanical metamaterials in Section \ref{subsection:shepp3d} and \ref{subsection:cubic}. 

\subsection{Numerical convergence}
\label{subsec:convergence}

\begin{figure*}
    \centering
    \includegraphics[width=\linewidth]{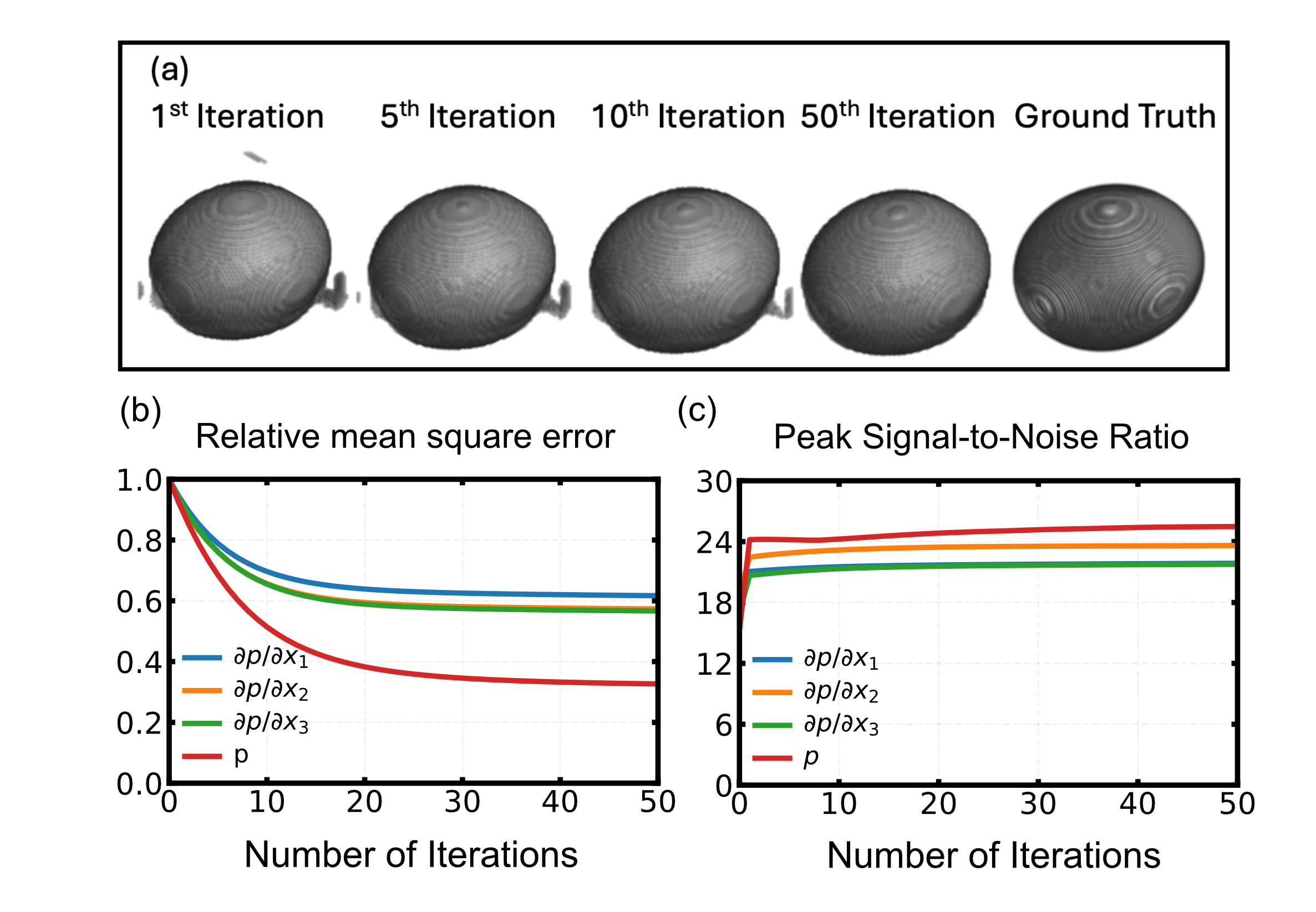}
    \caption{Convergence of the estimated pressure gradient fields and the pressure field (a) Reconstructed 3D morphology of the phantom after 1, 5, 10, and 50 iterations, shown alongside the ground truth. (b) The relative mean square error (RMER) as a function of the iteration numbers. (c) The peak signal-to-noise ratio  (PNSR) as a function of the iteration numbers. The ground truth is the \textit{Shepp3d} phantom with the size of $65^3$. The noise level is set to be zero.}
    \label{fig:convergence}
\end{figure*}
With the proposed setup described in Section~\ref{sec:physics}, we evaluated the performance of the algorithm introduced in Section~\ref{sec:inverse} using synthetic source data based on the \textit{Shepp3d} matrix of size \( 65^3 \) without adding noise. Detailed parameter settings are listed in Table~\ref{tab:parameters}. Note that, while using a translational stage with a single probe beam and sensor is cost-effective, it significantly increases acquisition time. For simplicity, the number of sensors was set to match the number of intervals: 65 sensors were placed along each edge for the \textit{Shepp3D} phantom and cubic fields, and 33 sensors for the mechanical metamaterials. In principle, one may employ a translational stage and use only a single probe beam and sensor, at the expense of longer acquisition time.

Figure~\ref{fig:convergence} illustrates the convergence of the the reconstruction. Figure~\ref{fig:convergence} (a) shows the 3D reconstructed morphology after the 1st, 5th, 10th, and 
50th iterations, while Figure~\ref{fig:convergence} (b) and (c) present the RMER and PSNR, respectively, 
as functions of number of iteration.  
The images in Figure~\ref{fig:convergence}(a) are obtained by post-processing the reconstructed pressure fields. Specifically, the volume was normalized and segmented using Otsu’s adaptive thresholding, with sub-threshold voxels suppressed to enhance high-intensity features.

As quantitatively shown in Figure~\ref{fig:convergence}(b), both pressure gradient fields and pressure fields exhibit rapid convergence during the 
first 20 iterations, followed by a slower refinement phase. After 50 iterations, the RMER for the pressure gradient field stabilizes around 60\%, while for the pressure field, it converges to approximately 32\%. Owing to the ill-posedness of the inverse problem, introducing $L_1$ regularization promotes convergence by stabilizing the solution, even when the sensor data are sparse. Interestingly, the prediction of the pressure field achieves a higher accuracy than that of the pressure gradient field,  because integration using  finite element methods utilizes more measurement information and reduces the impact of numerical error. In practical applications of source reconstruction, the primary goal is the relative similarity between predictions and ground truth. As a result, PSNR serves as a more effective metric. Fig.~\ref{fig:convergence}(c) plots the PSNR as a function of the number of iterations, demonstrating that even after the first iteration, the algorithm successfully reconstructs most of the subject's relative topology. The slower convergence of the RMER might be attributed to the fixed learning rate used in the ISTA formulation. In this study, all learning rates are set to be 0.1. For image reconstruction using the adjoint operator, this learning rate requires at least 10 iterations to reach convergence. A higher learning rate accelerates convergence but may cause instability.

\subsection{Algorithm demonstration for Shepp3D phantom}
\label{subsection:shepp3d}
\begin{figure*}
    \centering
    \includegraphics[width=0.9\linewidth]{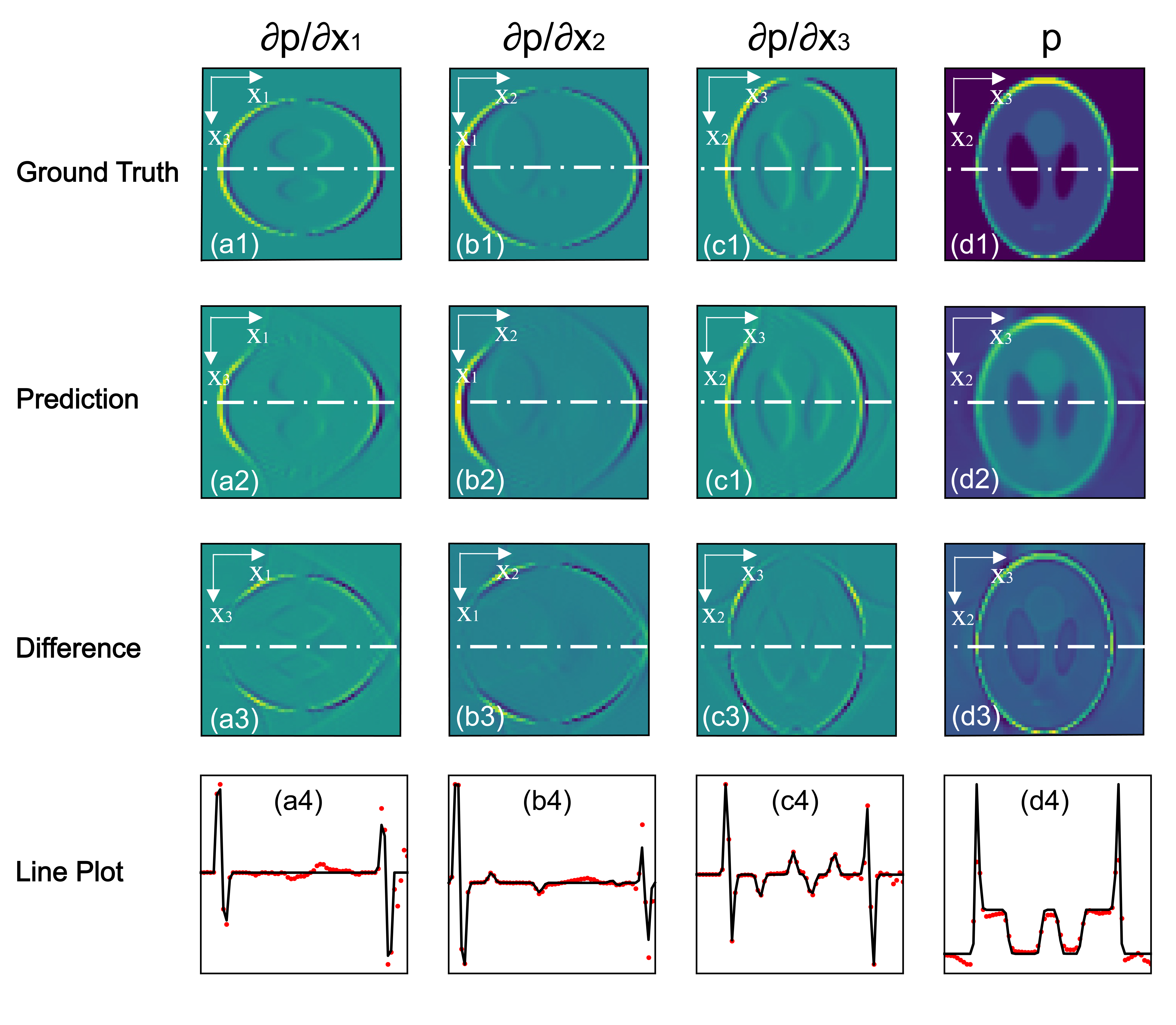}
    \caption{Comparisons of the ground truth and the predictions for the pressure gradient fields and pressure field. From left to right: (a1, a2, a3, a4): $\frac{\partial p}{\partial x_1}$;(b1, b2, b3, b4): $\frac{\partial p}{\partial x_2}$; (c1, c2, c3, c4): $\frac{\partial p}{\partial x_3}$; (d1, d2, d3, d4): $p$. From top to bottom: (a1, b1, c1, d1): ground truth; (a2, b2, c2, d2): predictions based on the image 0.1;  (a3, b3, c3, d3): Difference between the ground truth and the predictions; (a4, b4, c4, d4): The amplitude of the signal as a function of the distance along the white dash-dot line for both the ground truth (solid black line) and the prediction (red dotted line). The ground truth is the \textit{Shepp3d} phantom with the size of $65^3$. The predictions are obtained from the image 0.1 after 50 iterations and the finite element method. The reconstruction was assuming 65 probe beam positions and detectors separated by 1 mm, a laser duration of 10 ns and a recording duration of 100 $\mu$s. The Pearson correlation coefficients between the reconstructed signals and the ground truths in (a4), (b4), (c4), and (d4) are 96.43$\%$, 98.20$\%$, 99.45 $\%$ and 92.22 $\%$, respectively. 
 }
    \label{fig:shepp3d}
\end{figure*}

Given the convergence with regularization after 50 iterations, we visualized the ground truth, predicted results, and their differences at that point.
Figure ~\ref{fig:shepp3d} shows the ground truth with the predicted pressure gradient and pressure fields using both 2D and 1D slices with the number of iterations 50. The 2D slices were taken from the central plane, with the coordinates specified in the corresponding figures, while the 1D slices were extracted along the central line, indicated by the white dash-dot line in the 2D slices. As shown in Figure.~\ref{fig:shepp3d}(a, b, c), the prediction accuracy of the pressure gradient field is higher in the direction perpendicular to the derivative component. For sensors distributed across different planes, the different orientations help mitigate the issue of low resolution in alternative directions. This improves the accuracy of the predicted final pressure field, which, through integration by finite element methods, achieves higher resolution than any individual pressure gradient field alone. The convergence of RMSE and PSNR can be improved by increasing the number of sensors at the measurement planes.

\begin{figure*}
    \centering
    \includegraphics[width=0.9\linewidth]{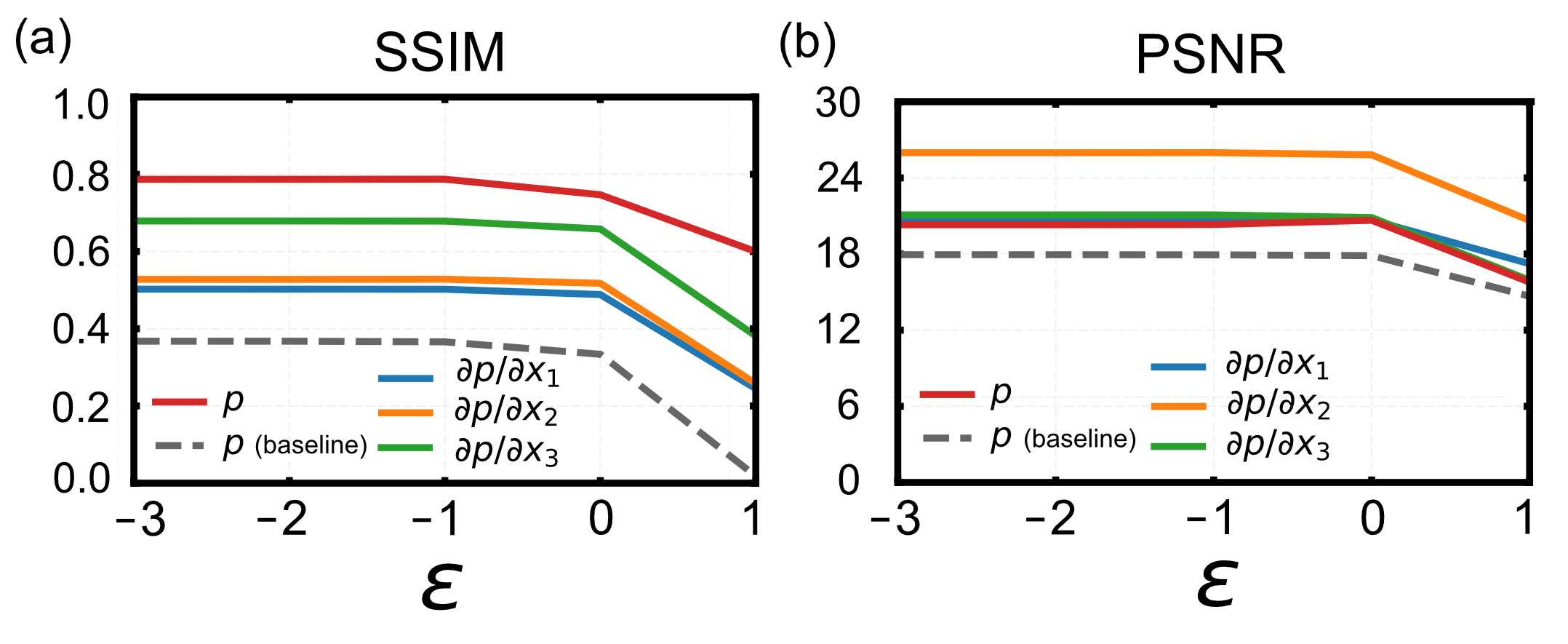}
    \caption{The new modality outperforms conventional method across different noise levels. (a) Structural similarity index measure (SSIM) and (b) peak signal-to-noise ratio (PSNR) over the relative error amplitude $\epsilon$, introduced as sensor measurement error during the Radon Transform step. The baseline conventional pressure-based approach uses a central frequency of 1 MHz, with a lower cutoff frequency of 0.2 MHz and an upper cutoff frequency of 2 MHz.}
    \label{fig:shepp3d-ssim}
\end{figure*}

Next, we investigate the impact of the sensor measurement error (i.e., the signals after the Radon transform) on the SSIM and PSNR of the pressure field reconstruction, as shown in Figure ~\ref{fig:shepp3d-ssim}. Each data point in the Radon transform image is perturbed by Gaussian noise, randomly sampled and scaled by the maximum absolute amplitude. For the conventional pressure-based approach used as a baseline, transducers were deployed on three faces of the domain with 33 × 33 elements per face. The results plotted were obtained under the same noise level (Fig. ~\ref{fig:shepp3d-ssim} dashed lines). To model the limited bandwidth of a conventional pressure transducer,   each transducer has a center frequency of 1 MHz and a bandwidth spanning 0.2–2.0 MHz. This band-limiting was implemented by applying a frequency mask to the fast Fourier transform of the pressure fields, followed by an inverse Fourier transform using MATLAB fft and ifft functions \cite{Rayleigh1945}. For the comparison, all simulations were conducted using a characteristic length reduced by a factor of 40 to highlight mesoscopic scale features.  Additive noise at varying levels was then introduced after filtering. As shown in Fig. 5, the proposed modality outperforms the pressure-gradient approach in both SSIM and PSNR. The image
reconstruction framework remained highly stable even at maximum error levels as high
as 1. This stability arises from the unbiased nature of the Gaussian distributed errors,
which are mitigated during the minimum potential energy optimization inherent in the
Galerkin formulation. Consequently, even at elevated noise levels, the finite-element reconstruction preferentially extracts the components of the gradient data consistent with the governing Poisson equation, effectively attenuating random fluctuations. Therefore, the pressure-gradient–based approach exhibits superior stability in noisy measurement environments.

\subsection{Algorithm demonstration for cubic field and mechanical metamaterials}
\label{subsection:cubic}

\begin{figure*}
    \centering
    \includegraphics[width=0.9\linewidth]{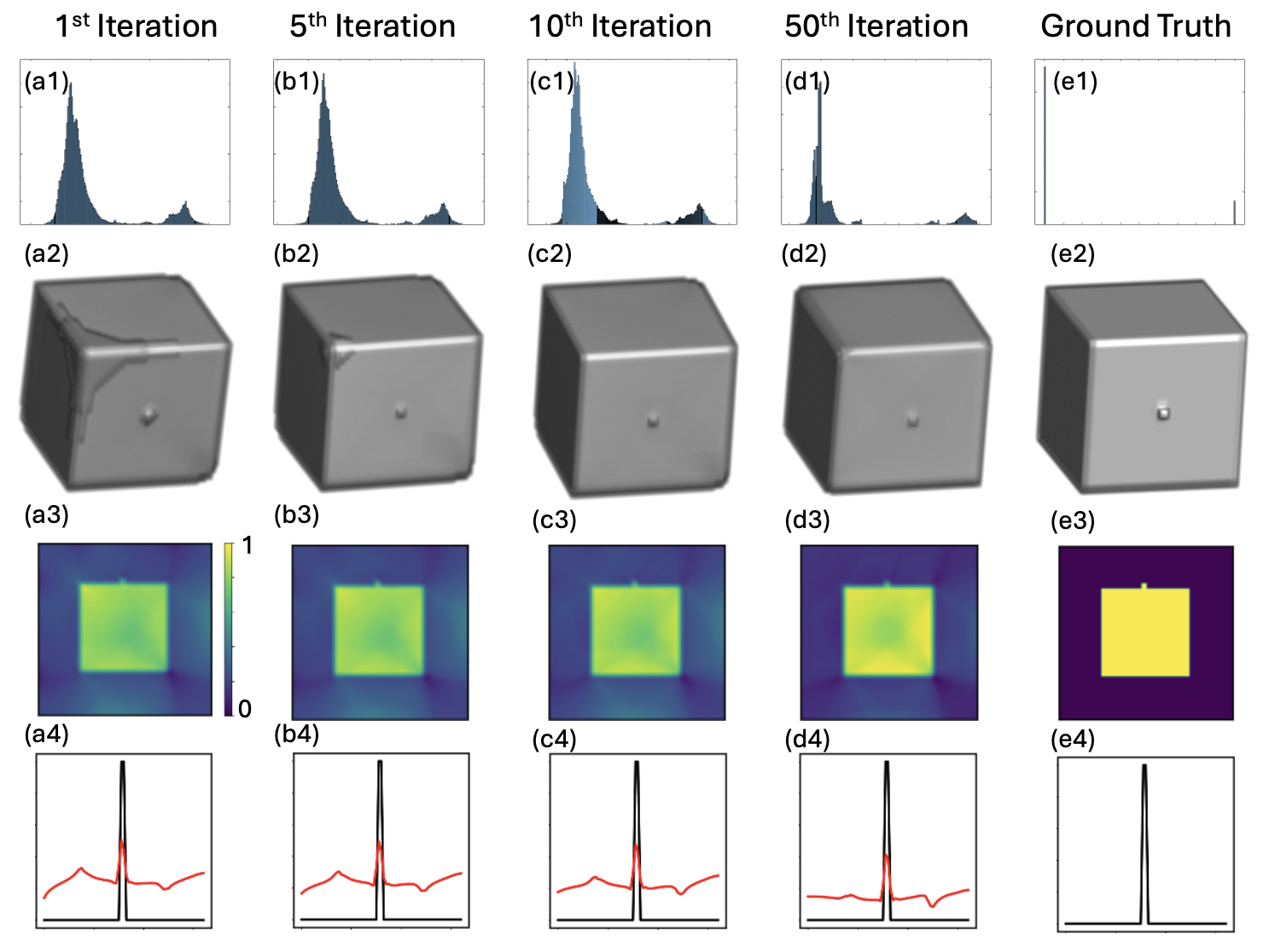}
    
\caption{Reconstruction results for the internal cube with a small bulge. Each column corresponds to different number of iteration (a) 1,  (b) 5,  (c) 10, (d) 50 and is compared against (e) the ground truth. Each row shows: (1) the normalized histogram of the initial pressure-intensity values, representing the statistical distribution used for the subsequent Otsu's thresholding; (2) the reconstructed 3D morphology; (3) a representative cross-sectional view across the small bulge of the normalized pressure-intensity contour (color bar in (a3) uses the normalized pressure scale); and  (4) a representative cutline profile across the small bulge of the normalized pressure intensity (red: prediction; black: ground truth). }
\label{fig:cubic}
\end{figure*}

To further evaluate the algorithm’s ability to reconstruct small, 
tumor-like features, we introduce a cubic phantom containing sharp 
discontinuities with a tiny bulge (Figure~\ref{fig:cubic} (e2)). 
Such fine-scale structures are well known to pose challenges for numerical reconstruction algorithms.

\begin{figure*}
    \centering
    \includegraphics[width=0.55\linewidth]{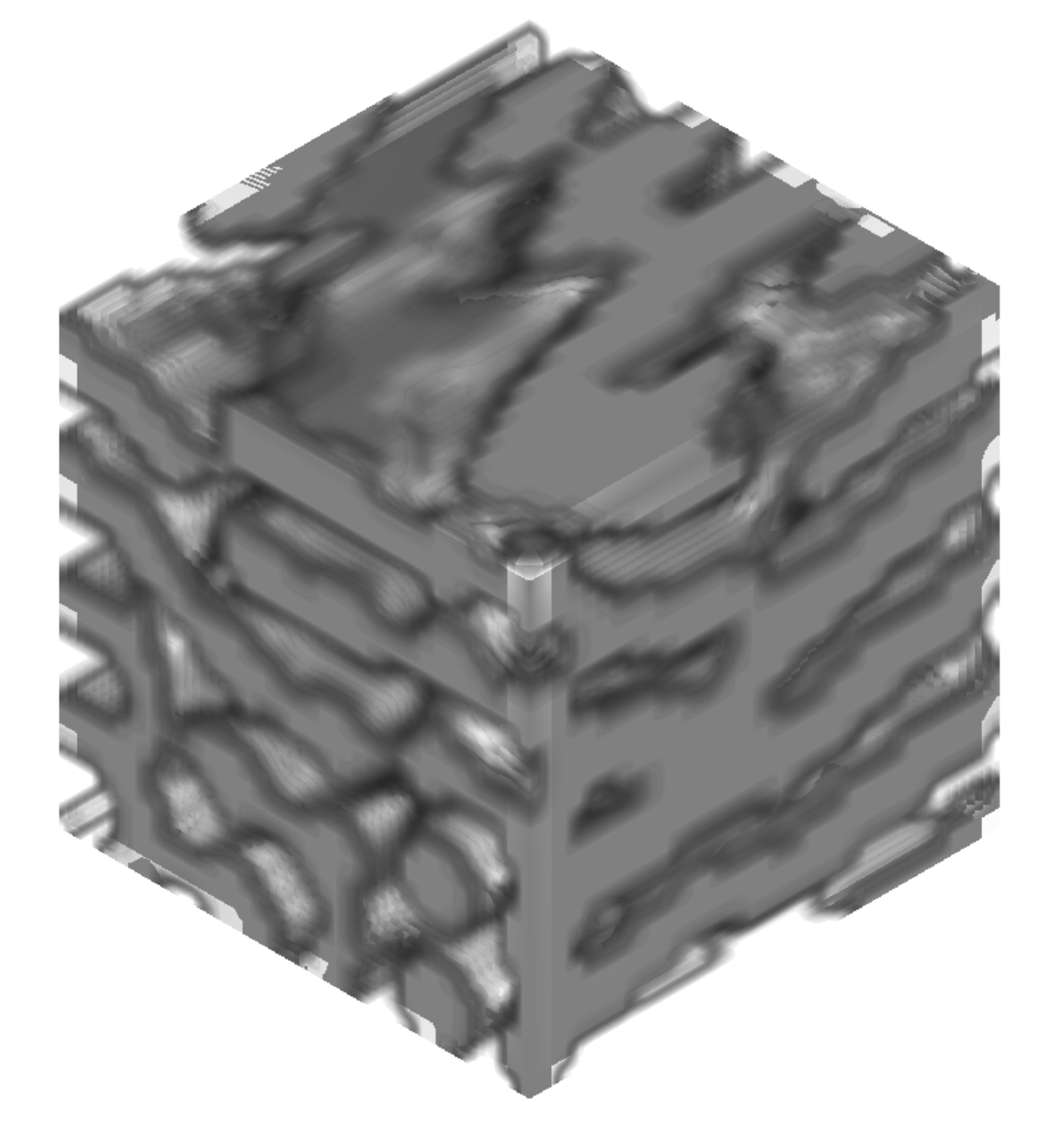}
    \caption{Representative structure of a lamellar-type spinodal metamaterial of size $32^3$. 
}
\label{fig:spinodal_source}
\end{figure*}

\begin{figure*}
    \centering
    \includegraphics[width=0.9\linewidth]{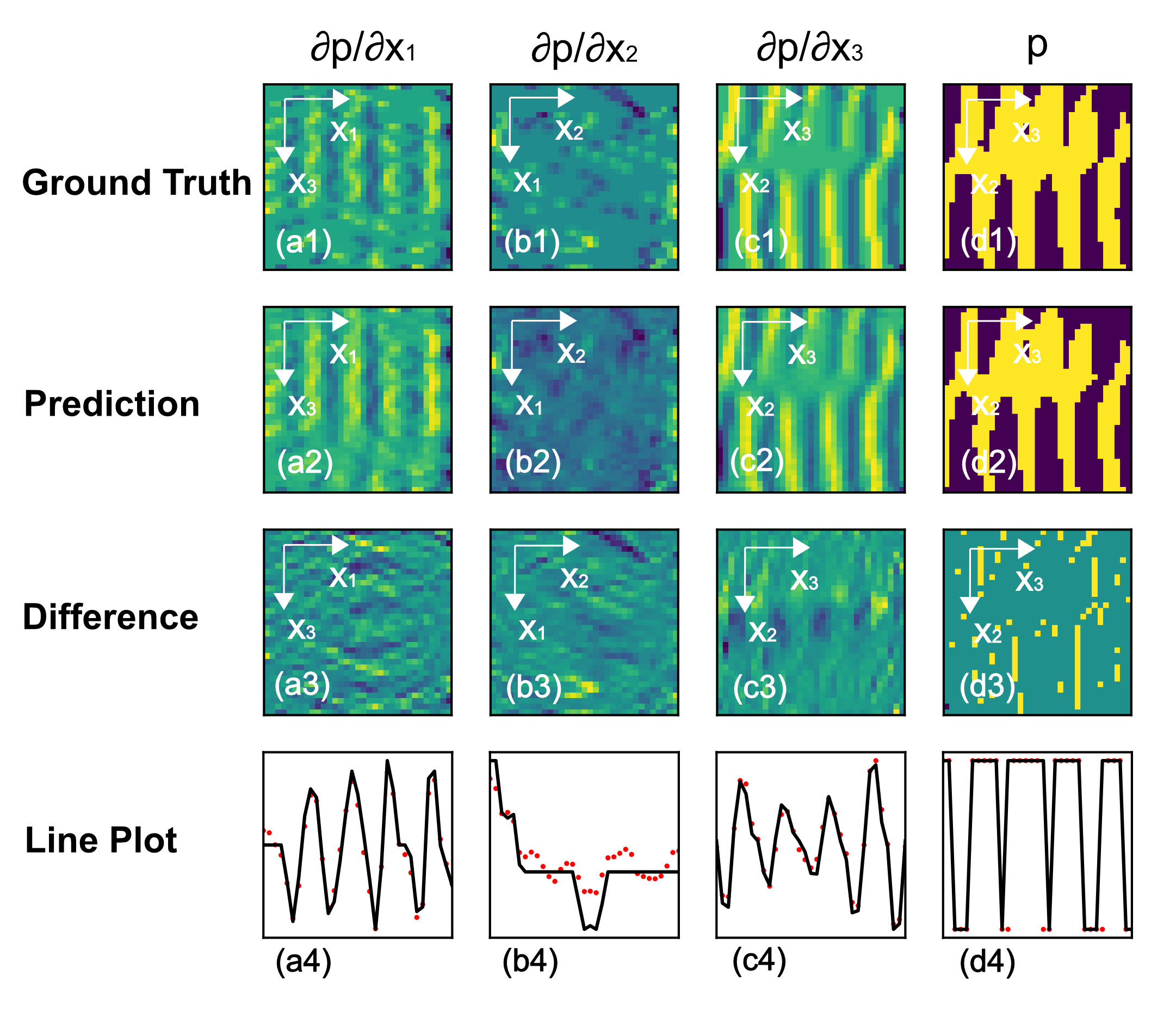}
    \caption{For the data source of the spinodal metamaterials, comparisons of the ground truth and the predictions for the pressure gradient fields and pressure field. (a1, a2, a3, a4) $\frac{\partial p}{\partial x_1}$; (b1, b2, b3, b4) $\frac{\partial p}{\partial x_2}$; (c1, c2, c3, c4) $\frac{\partial p}{\partial x_3}$; (d1, d2, d3, d4) $p$; (a1, b1, c1, d1) ground truth; (a2, b2, c2, d2) predictions based on the image reconstruction;  (a3, b3, c3, d3) Difference between the ground truth and the predictions; (a4, b4, c4, d4) The amplitude of the signal as a function of the distance along the white dash-dot line for both the ground truth (solid black line) and the prediction (red dotted line). The ground truth is the cubic source with the size of $32^3$. The predictions are obtained from the image reconstruction after 50 iterations and the finite element method. The reconstruction was assuming 33 probe beam positions and detectors separated by 1 mm, a laser duration of 10 ns and a recording duration of 100 $\mu$s. The Pearson correlation coefficients between the reconstructed signals and the ground truths in (a4), (b4), (c4), and (d4) are 98.48 $\%$, 93.26$\%$, 98.87 $\%$ and 97.22 $\%$, respectively. 
     }
\label{fig:spinodal}
\end{figure*}
Figure~\ref{fig:cubic} presents the ground-truth with  a pulse-like profile along the edge of the structure and reconstruction results for a cubic field containing a tiny bulge. The comparison between the predicted and true distributions (Figure~\ref{fig:cubic} (a2–e2, a3–e3)) demonstrates that the method accurately recovers the locations of the sharp edges of the cube and the small bulge, enabling a faithful reconstruction of the overall morphology.  
The amplitude deviations in the small feature (Figure~\ref{fig:cubic} (a4–d4)) arise from different numerical steps, which are affected by the number of sensors and the discretization mesh size.
Overall, these results highlight the method’s ability to recover sharp spatial features and detect small structures that may be overshadowed by larger components—capabilities that are particularly important for small tumor detection and for characterizing microscale mechanical metamaterials, where dynamic and non-destructive assessment remains a challenge \cite{diebold1990photoacoustic, calasso2001photoacoustic, kai2023dynamic, jin2025characterization}.

Microscale mechanical metamaterials offer unique functionalities but are typically characterized using destructive techniques. Here, we simulated the optical detection of laser-induced vibrations \cite{bai2023ponderomotive} in mechanical metamaterials and applied the reconstruction algorithm to recover the underlying structure. We introduced a lamellar-type spinodal metamaterial of source size $33^3$ as shown in Figure \ref{fig:spinodal_source}, which were selected due to their exceptional mechanical performance and bioinspired architecture \cite{portela2020extreme}. It enables precise modulation of mechanical anisotropy, making these materials particularly well-suited for investigating the relationship between geometry and mechanical function in complex systems, as well as for designing architected materials with programmable, application-specific responses \cite{xia2022responsive}. The details of forward simulations and inversion are in Sections \ref{sec:physics} and \ref{sec:inverse}. Figure \ref{fig:spinodal} shows the reconstruction of the photoacoustically excited spinodal metamaterials. 
As shown in Figure \ref{fig:spinodal} (d4), the method accurately captures the lamellar-like mechanical characteristic structure, highlighting the potential of the proposed method to resolve the characteristic signatures of microscale mechanical metamaterials \cite{jin2023recent, kai2023dynamic}.

\section{Conclusion}

Photoacoustic tomography enables imaging with deep acoustic penetration depth but suffers from resolution or sensitivity issues due to the limitations of acoustic transducers. We proposed a new photoacoustic imaging modality by recording acoustic fields using optical beam deflections. An optimization-based reconstruction was proposed to reconstruct the initial pressure gradient field, and a Galerkin method was used to reconstruct the pressure field from the pressure gradient field. Simulations showed excellent reconstruction performance for simple sources, Shepp-Logan phantoms, and mechanical metamaterials. 

The key advances described in this paper include: (1) delineation of a new imaging modality based on optical deflection measurements of the photoacoustic waves; (2) modeling the multiphysics processes underlying the forward measurements; (3) the use of a robust optimization-based method for solving the inverse problem in three dimensions; (4) a detailed performance evaluation and benchmark determined over a wide range of imaging setups. 

However, limitations also exist. As with other photoacoustic methods, the modality still lacks the spatial resolution at the nanoscale necessary to resolve molecular-scale interactions\cite{yan2025identifying}. The numerical method encounters the conventional challenge of model-based optimization — high computational cost, especially wave dynamics simulations in three dimensions \cite{benning2018modern, cox2007k}. Furthermore, accurately integrating the pressure field necessitates knowledge of all three pressure gradient components, which in turn requires a dense arrangement of sensor placement on three non-parallel planes or an extended data acquisition period. Finer grids were prevented from satisfying the memory and computational constraints. Alternative spatial arrangements of sensors and their associated properties remain unexplored. 

In addition, other heterogeneous properties of the acoustic medium can also be modeled and explored in the inverse problem set-up \cite{treeby2012modeling}. Recent LED-based photoacoustic imaging systems focus on low-cost and portable optical excitation, offering promising routes toward clinical translation. In contrast, the proposed modality addresses a complementary bottleneck by enabling all-optical ultrasound detection, and can be naturally combined with LED-based excitation to provide enhanced miniaturization, and greater system integration \cite{zhu2018light, francis2020tomographic, zhu2020towards, hariri2018characterization}.

Note that linear Radon transforms holds in Eq. \ref{radon_formula} when the refractive index gradients are mild and ray bending is limited. For media with stronger heterogeneity, more accurate modeling—such as optical path simulations based on ray-propagation methods (e.g., wavefront tracking)—would be required. Future work may also explore advanced approaches, including deep neural networks, to address multiphysics inverse problems \cite{raissi2019physics, antholzer2019deep, bar2019learning}.

This paper focuses on developing physical models and inversion algorithms tailored to the newly proposed imaging modality. The claims regarding the superiority of the all-optical photoacoustic tomography require further validation through experimental research. It remains to be demonstrated whether the newly proposed modality yields a superior signal-to-noise ratio, as well as improved resolution and contrast in the reconstructed images, compared to results obtained using conventional pressure-based detection methods. In the experimental realization of the proposed setup, a simplified device employing a simple probe laser and detector can be scanned along the length of the boundary. Both the probe laser and the detectors will be mounted on a translation stage that moves along the detector coordinate as the pulsed laser fires while the time dependence of the deflection signal is recorded. In the follow up experiments, the determination of best sources, detectors and optical trains can be carried out with this single channel instrument.

\section*{Acknowledgment}
X.Y. is supported by NSF-Simons Center for the Mathematical $\&$ Statistical Analysis of Biology (DMS-174269) and the
Harvard Quantitative Biology Initiative. The computations in this paper were conducted using computational
resources and services at the Center for Computation and Visualization, Brown University. We thank Prof. Gerald J. Diebold at Brown University and Dr. Zhi Li at Nanyang Technological University for insightful discussions and proof-reading of the manuscript.

\newpage
\bibliography{apssamp} 
\end{document}